\documentclass{WileyMSP-template}

\usepackage{amsmath}

\begin{document}

\pagestyle{fancy}
\rhead{\includegraphics[width=2.5cm]{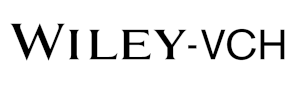}}

\title{Adsorption of water molecules on pristine and defective NiPX$_3$ (X: S, Se) monolayers}

\maketitle

\bigskip
\author{Zhicheng Wu},
\author{Sifan Xu},
\author{Yong Zhou},
\author{Qilin Guo},
\author{Yuriy Dedkov$^*$},
\author{Elena Voloshina$^*$}

\bigskip
\begin{affiliations}
Z. Wu, S. Xu, Y. Zhou, Q. Guo, Prof. Y. Dedkov, Prof. E. Voloshina\\
Department of Physics, Shanghai University, Shangda Road 99, 200444 Shanghai, China\\
Email Address: yuriy.dedkov@icloud.com; elena.voloshina@icloud.com

\medskip
Prof. Y. Dedkov, Prof. E. Voloshina\\
Centre of Excellence ENSEMBLE3 Sp. z o.\,o., ul. Wolczynska 133, 01-919 Warsaw, Poland\\
Institut f\"ur Chemie und Biochemie, Freie Universit\"at Berlin, Arnimallee 22, 14195 Berlin, Germany
\end{affiliations}


\begin{abstract}
Layered transition metal trichalcogenides MPX$_3$ (M: transition metal; X: S, Se) demonstrate a wide spectrum of properties and are widely proposed as effective materials for the water splitting reactions. Among these materials, NiPX$_3$ are the most promising ones due to the match their electronic structures for the oxygen and hydrogen evolution reactions. Here, we present first steps of a detailed theoretical description on the adsorption of water molecules on pristine and defected (chalcogen vacancies) surfaces of NiPX$_3$ and show that in all cases a physisorption takes the place with adsorption energies do not exceeding $-650$\,meV and water dissociative adsorption is unfavourable. This work provides a general description for water molecules interaction with MPX$_3$ and can serve as a basis for further studies on more complicated water/MPX$_3$ reactions.
\end{abstract}

\section{Introduction}
\label{S:Intro}

Metal phosphorus trichalcogenides (MPX$_3$ with  M: transition metal and X: S, Se) are the van der Waals (vdW) layered crystalline materials, where each monolayer adopts a $D_{3d}$ point symmetry (Fig.~\ref{fig:structure}a,b). Six M cations form a honeycomb lattice with a phosphorus dimer which perpendicularly crosses the centre. The P\,--\,P dimers are covalently bound to six sulphur (or selenium) atoms to form an ethane-like (P$_2$X$_6$)$^{4-}$ unit, where each P-atom is tetrahedrally coordinated with three S (or Se) atoms. Meanwhile, each sulphur (or selenium) atom is coordinated with M sites and is covalently bonded to one P atom. The two-dimensional (2D) layers are identical across the MPX$_3$ family members, but their 3D stacking arrangement depends on cations and anions constituting the crystals. Most of them crystallise into monoclinic ($C2/m$)~\cite{Ouvrard:1985hi} or trigonal ($R\bar 3$)~\cite{Wiedenmann:1981bs} crystallographic structures.

The unique crystal structure of MPX$_3$ with wide vdW gaps between the single layers and strong ionic bond between M cations and (P$_2$X$_6$)$^{4-}$ endows these materials with rich functionalities. For example, the comparatively large band gap ranging from $1.3$\,eV to $3.5$\,eV,~\cite{Du:2016ft,Zhang:2016kr} which is unavailable for other 2D materials, indicates their potential application in optoelectronics and photo(electro)chemical catalysis~\cite{Wang:2018dh}.

In consequence of the small cleavage energy, MPX$_3$ crystals can be easily exfoliated into 2D honeycomb monolayers~\cite{Du:2016ft}, making them attractive for the 2D materials community. Here, one of the research directions is motivated by potential application of MPX$_3$ in low-dimensional magnetic and spintronic devices~\cite{Li:2014de,Joy:1992cf,Goossens:2011kl,Rao:1991bj} and the detailed understanding of the electronic and magnetic properties of MPX$_3$ monolayers is expected from experimental and theoretical fundamental studies. 

Until now the vast majority of the relevant works are theoretical calculations~\cite{Zhang:2016kr,Chittari:2016cd,Gu:2019kw,Yang:2020ex}, whereas the experimental studies on magnetic properties are mainly focused on the MPX$_3$ bulk crystals~\cite{Wiedenmann:1981bs,Joy:1992cf,Takano:2004km,Wildes:2015iv,Wildes:2017ew,Brec:1980ts,Lancon:2018gf,Dedkov:2020ca}.  According to density functional theory (DFT) calculations, the 2D MPX$_3$ exhibit a variety of magnetic behaviours~\cite{Sivadas:2015gq}, including ferromagnetism (FM), N\'eel anti-ferromagnetism (nAFM), zig-zag anti-ferromagnetism (zAFM), and stripy anti-ferromagnetism (sAFM), which can be modulated through doping~\cite{Li:2014de,Pei:2018kb,Yang:2020ko} or lattice strain effects~\cite{Chittari:2016cd}. Surprisingly, the role of chalcogen vacancies, which may appear during exfoliation and processing was not taken into account. This kind of defects may deteriorate the performance of future MPX$_3$-based devices. At the same time, deviations from perfection can be useful in some applications, as they make it possible to tailor the local properties of material and to achieve new functionalities.

Presently, the focus in the studies of this class of materials is shifted to the applied-oriented investigations, while there are no basic and systematic studies on their electronic structure, which provide direct insight in the properties, which make MPX$_3$ family so promising candidates for different applications. For example, despite attempts to apply these materials for the water splitting reactions~\cite{Wang:2018dh,Gusmao:2017kz}, the fundamental studies on the interaction of transition-metal phosphorus trichalcogenides with a single water molecule are missed. The aim of our studies is to fill this lacuna.

The main objects of this work are the typical representatives of the family of transition-metal phosphorus trichalcogenides -- NiPS$_3$ and NiPSe$_3$. We start with the description of the electronic and magnetic properties of 2D NiPX$_3$ (X = S, Se). Here, the chalcogen defects (Fig.~\ref{fig:structure}c-e) are addressed. Since NiPX$_3$ (among other MPX$_3$ materials) were proposed for the possible application in water-splitting reactions~\cite{Jenjeti:2016hz,Li:2019cz}, on a first step, we performed detailed theoretical studies of water molecules adsorption on pristine and defected (S- and Se-vacancies) surface of NiPX$_3$. Also, these modelings are important taking in mind the future utilisation of 2D NiPX$_3$ materials in real low-dimensional applications, where adsorption of H$_2$O molecules can drastically influence the electronic structure of 2D NiPX$_3$ (X = S, Se) and thus functionality of devices. [Note: The present study deals with the adsorption of H$_2$O single molecules on NiPX$_3$ and is not devoted to the consideration of any electrochemical water splitting reactions.]

\section{Results and Discussion}

Bulk NiPS$_3$ crystallises in the $C2/m$ space group, while NiPSe$_3$ in $R\bar3$ and both of them adopt the AFM ground state. (For optimised 3D structures and calculated ground-state properties, see Supplementary Information, Tab.\,S1 and Fig.\,S1). It is expected that individual NiPX$_3$ layers can be isolated by mechanical exfoliation. In this regard, the cleavage energy ($E_\mathrm{cl}$) of a layered material is an essential property that need to be considered. The calculated values, $E_\mathrm{cl}(\textrm{NiPS}_3)=0.19$\,J\,m$^{-2}$ and $E_\mathrm{cl}(\textrm{NiPSe}_3)=0.26$\,J\,m$^{-2}$ (see Supplementary Information, Fig.\,S2), are smaller than that of a  bulk graphite ($0.36$\,J\,m$^{-2}$) \cite{Zacharia:2004go}, which is used as an indicator for the feasibility of exfoliation of materials in experiments. It is also important to note that the obtained values agree with the previously published data for NiPS$_3$~\cite{Du:2016ft}.

\begin{figure}[h]
\centering
 \includegraphics[width=0.92\textwidth]{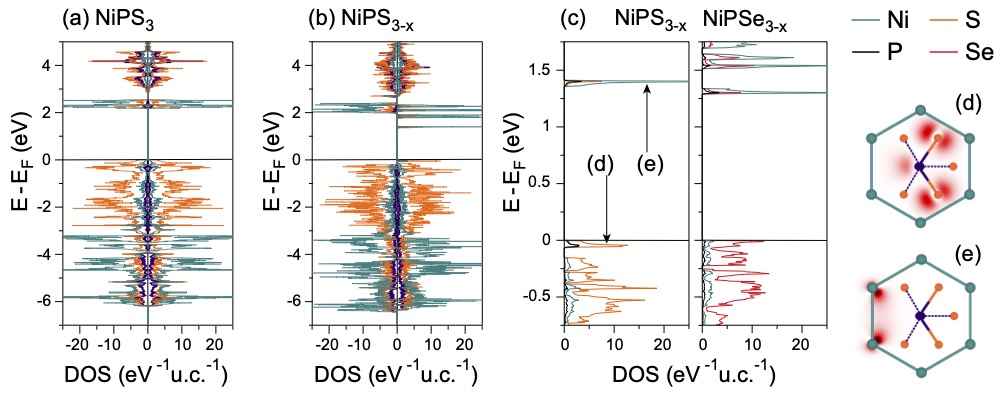}
 \caption{Site-projected density of states for (a) pristine NiPS$_3$ and (b) defective NiPS$_3$.  (c)  The zoomed image of the density of states in the vicinity of $E_F$ for the defective NiPS$_3$ (left) and NiPSe$_3$ (right). (d,e) Top view of the relaxed structure of defective NiPS$_3$ superimposed with the electron densities calculated for the occupied and unoccupied states marked respectively at (c). Green, violet, and orange spheres  represent Ni, P, and S atoms, respectively. }
 \label{fig:dos}
\end{figure}

When studying 2D NiPX$_3$ monolayers, we consider four different magnetic configurations of Ni$^{2+}$ ions, which include FM, nAFM, zAFM, and sAFM configurations, as well as the non-magnetic states. According to our calculations, the energetically most favourable structure corresponds to the zAFM arrangement of magnetic moments for the both NiPS$_3$ and NiPSe$_3$ (Tab.~\ref{tab:2D}). The both monolayers under study are  indirect band gaps semiconductors: the VBM is located at the K point, while the CBM lies on the K to $\Gamma$ path (see Supplementary Information, Fig.\,S3). Moreover, the top of the valence band is mainly formed by S/Se, whereas the bottom of the conduction band consists of S/Se and Ni. These observations are consistent with the data recently published for bulk NiPS$_3$, which was shown to be a charge-transfer insulator~\cite{Yan:2021cn}. The band gaps calculated by means of PBE$+U$+D2 are $E_g=2.19$\,eV and $1.85$\,eV for NiPS$_3$ and NiPSe$_3$, respectively (see Fig.~\ref{fig:dos}a and Supplementary Information, Fig.\,S3). Hybrid functional (here: HSE06) yields qualitatively similar results, albeit yielding wider gaps: $2.87$\,eV and $2.46$\,eV for 2D NiPS$_3$ and NiPSe$_3$ (see Supplementary Information, Fig.\,S3), which are consistent with previously published data ($3.01$\,eV for NiPS$_3$~\cite{Zhang:2016kr}). 

The calculated total energies for different magnetic configurations were used to estimate the exchange-coupling parameters ($J_1$, $J_2$, and $J_3$), describing the magnetic interactions between Ni$^{2+}$ ions (Tab.~\ref{tab:J}). The obtained results can be understood on the basis of the Goodenough-Kanamori-Anderson (GKA) rules~\cite{Goodenough:1955th,Kanamori:1960ki}. Two types of interactions have to be considered: Direct Ni--Ni couplings ($t_{2g}-t_{2g}$ orbital overlapping) and indirect Ni--X--Ni super-exchange couplings. The $t_{2g}$ orbitals are filled in NiPX$_3$ and direct couplings do not exist. The super-exchange interaction between nearest neighbours is through the two X atoms on edge-shared octahedra between neighbouring Ni atoms. The respective Ni--X--Ni angle is $85.3^\circ$ for X = S and $86.3^\circ$ for X = Se, i.\,e. close to $90^\circ$. Thus, according to the GKA rules, the considered interaction should be ferromagnetic ($J_1>0$). In the case of second neighbour coupling, no coplanar X atoms with overlapping orbitals exist, which results in $J_2\approx 0$. The super-super-exchange interaction involves a Ni--X--X--Ni bridge with two X atoms belonging to the same PX$_3$ sublayer. Thus, the third neighbour interaction is antiferromagnetic ($J_3<0$). Overall, the zAFM ground state is determined by 3NN super-super-exchange interaction. Our results are in good agreement with the previously published experimental data for NiPS$_3$~\cite{Lancon:2018gf}. 

The calculated $J$ parameters were used in the Monte-Carlo simulations of N\'eel temperatures of NiPX$_3$ monolayers. The estimated $T_N$ is $148$\,K for 2D NiPS$_3$ and $183$\,K for 2D NiPSe$_3$ (see Supplementary Information, Fig.\,S4), which are close to the available experimental results ($T_N$ is $155$\,K~\cite{Lancon:2018gf} and $206$\,K~\cite{LeFlem:1982gs} for NiPS$_3$ and NiPSe$_3$, respectively).

\begin{figure}[h]
\centering
 \includegraphics[width=0.48\textwidth]{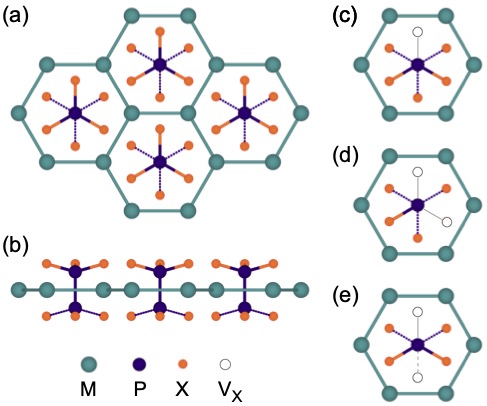}
 \caption{(left panel) Top (a) and side (b) views of a single layer of MPX$_3$. (right panel) Structure of the considered chalcogen defects: (c) V$_\mathrm{X}$@1L; (d) V$_\mathrm{X2}$@1L; (e) V$_\mathrm{X2}$@2L. Spheres of different size/colour represent ions of different type. }
 \label{fig:structure}
\end{figure}

Mechanical exfoliation, usually utilised for monolayers production, can lead to formation of chalcogen vacancies. Therefore, we consider three different kinds of defects: (i) one vacancy at X-site, named as V$_\mathrm{S}$@1L or V$_\mathrm{Se}$@1L with a defect concentration about $2.5$\,\% (Fig.~\ref{fig:structure}c); (ii) two vacancies at the neighbouring X-sites of the same chalcogen sub-layer named as V$_\mathrm{S2}$@1L or V$_\mathrm{Se2}$@1L with a concentration about $5$\,\% (Fig.~\ref{fig:structure}d); (iii) two vacancies at X-sites of the different chalcogen sub-layers named as V$_\mathrm{S2}$@2L or V$_\mathrm{Se2}$@2L with a concentration about $5$\,\% (Fig.~\ref{fig:structure}e). 

The defect formation energies ($\Delta E_\mathrm{def}$) are presented in Table~\ref{tab:df}. Overall, $\Delta E_\mathrm{def}$ are in the same range that those of transition metal dichlcogenides -- sulphur vacancies in e.\,g. FeS$_2$ ($\Delta E_\mathrm{def}=2.25$\,eV~\cite{Tian:2020de}) and MoS$_2$ ($2.12$\,eV~\cite{Hong:2015if}),  and of reducible oxides -- oxygen vacancies in e.\,g. Fe-oxides ($\Delta E_\mathrm{def} = 1.72 \dots 3.51$\,eV \cite{Ovcharenko:2016fu,Mulakaluri:2010vy}). For the same defect type, Se vacancy is more likely to occur compared to S vacancy, which correlates with the respective electronegativity values. For the same X, the difficulty of defect formation is $\mathrm{V}_{\mathrm{X}2}@\mathrm{1L} > \mathrm{V}_{\mathrm{X}2}@\mathrm{2L} > \mathrm{V}_{\mathrm{X}}@\mathrm{1L}$ and from now on we will focus on detailed consideration of V$_\mathrm{X}$@1L.

A chalcogen vacancy formation leads to a strong modification of the local crystallographic and electronic structures (Fig.~\ref{fig:dos}). Thus, the phosphorus dimer is shifted from its original position. Notably, it is tilted towards the vacancy and its angle with the vertical direction is $2.8$ or $4.3$ degrees for X = S or Se, respectively. Furthermore, this dimer is pulled out from the the entire layer, while keeping the P--P distance almost unchanged with respect to the pristine structure. All P--X distances are slightly elongated and this effect is more pronounced for the defective sublayer (Tab.~\ref{tab:df}).

Upon removal of a chalcogen atom, the left behind electrons occupy the easily available electronic states of a (P$_2$X$_5$) entity. As a result, in the calculated density of states (DOS) of the NiPS$_{3-x}$ monolayer a localised defect state appears in the energy gap just below $E_F$ (Fig.~\ref{fig:dos}b,c) and the electron density is delocalised between the P and S atoms of the defective sublayer (Fig.~\ref{fig:dos}c,d). One more state is formed in the energy gap above $E_F$ and it has Ni-$3d$ character (Fig.~\ref{fig:dos}c,e). The magnetic moments of the Ni$^{2+}$-ions nearby the vacancy are coupled ferromagnetically and the both considered states appear in the spin-up channel. In the case of X = Se the occupied state is not well-localised, although its signature can be visible in the vicinity of $E_F$ (Fig.~\ref{fig:dos}c). Naturally, these modifications yield a decrease of the energy gap width: $E_g = 1.36$\,eV and $1.29$\,eV for the defective NiPS$_3$ and NiPSe$_3$, respectively. 

\begin{figure}[h]
\centering
\includegraphics[width=0.92\textwidth]{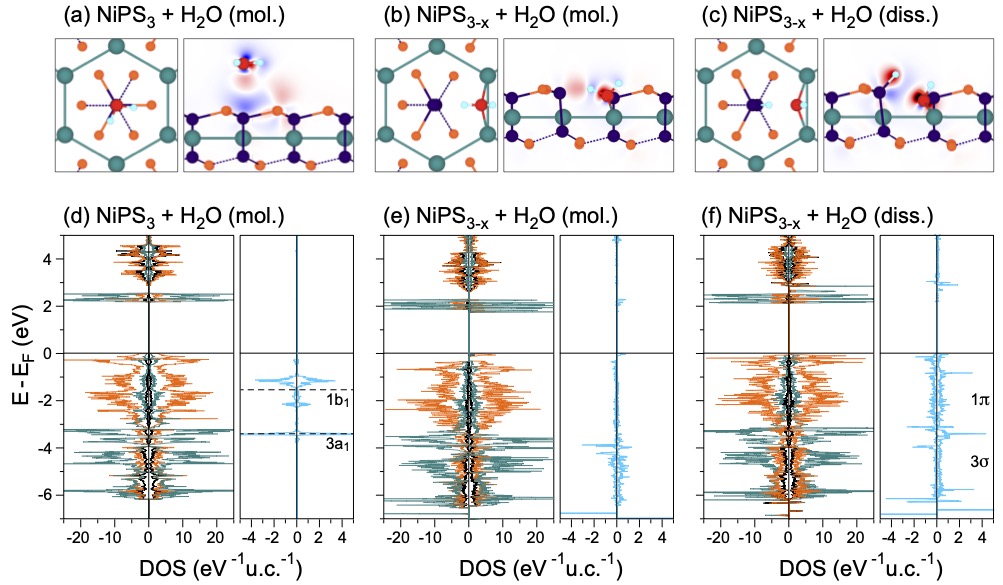}
 \caption{(a-c) Top and side views of the relaxed structures obtained after water adsorption on (a) pristine NiPS$_3$ (molecular adsorption); (b) defective NiPS$_3$ (molecular adsorption); (c) defective NiPS$_3$ (dissociative adsorption). Side view are superimposed with electron density redistribution maps. Electron density accumulation (depletion) is shown in red (blue). Green, violet, and orange spheres  represent Ni, P, and S atoms, respectively. The water molecule is shown with red and light-blue spheres, for O and H, respectively.
(d-f) Site-projected density of states after water adsorption for (d) pristine NiPS$_3$ (molecular adsorption); (e) defective NiPS$_3$ (molecular adsorption); (f) defective NiPS$_3$ (dissociative adsorption). The molecular orbitals of a gas-phase H$_2$O molecule are indicated by horizontal dashed lines. These are obtained by aligning the $2a_1$ core level of the gas-phase and adsorbed H$_2$O molecules.  }
 \label{fig:h2o}
\end{figure}

In the next step we investigate the adsorption of a single H$_2$O molecule on pristine NiPX$_3$ (X = S, Se) monolayers. Various high-symmetry adsorption sites and adsorption orientations are considered (see Supplementary Information, Tab.\,S2 and Fig.\,S5). All adsorption configurations have similar adsorption energies which range form $-115$\,meV to $-173$\,meV and from $-109$\,meV to $-176$\,meV for NiPS$_3$ and NiPSe$_3$, respectively. In the most stable adsorption structures, 
the oxygen atom of a water molecule stays directly above of 
an P atom [$d(\mathrm{P-O}) = 3.17$\,\AA\ and $3.34$\,\AA\ for NiPS$_3$ and NiPSe$_3$, respectively] and the H atoms are directed towards the neighbouring X atoms as it is shown in Figure~\ref{fig:h2o}a. In accordance with the weak interaction, the structural parameters of H$_2$O as well as of the studied monolayer undergo insignificant changes. 

To shed more light on the adsorbate mode, we have investigated the electron density redistribution upon adsorption ($\Delta\rho$). The $\Delta\rho$ and DOS plots are shown in Figure~\ref{fig:h2o}a,d. When molecular water adsorbs on the pristine NiPX$_3$ monolayer, the main charge rearrangement takes place between O and P atoms. Charge accumulation in the $p$-orbital of P on the side of the adsorbate indicates that it is the main orbital participating in the bonding. The electron density accumulation between H$_2$O and NiPX$_3$ is accompanied with a depletion at the hydrogen positions. Furthermore, there is some signature for the interaction between water hydrogens and the neighbouring chalcogen atoms. 

The main effects in DOS expected due to the interaction between H$_2$O and NiPX$_3$ are a shift of position and change in width of the molecular levels of the adsorbate. Analysis of the partial DOS (Fig.~\ref{fig:h2o}d) shows that the bonding to the monolayer is mainly due to the $1b_1$ molecular orbital (MO) of H$_2$O (the water lone pair). This orbital undergoes some broadening due to hybridisation with $p$ states of P. In addition, a slight upward shift with respect to the MO of gas phase water molecule is observed for this MO. (For the DOS of H$_2$O/NiPSe$_3$, see Supplementary Information, Fig.\,S6). 

The chalcogen defect binds the H$_2$O molecule more strongly: $E_\mathrm{ads}=-572$\,meV and $-637$\,meV for X = S and Se, respectively. This is related to the higher coordination of the molecule in the vacancy than on the pristine monolayer. The water molecule is coordinated between two Ni ions as it is shown in Figure~\ref{fig:h2o}b. Additional stability is due to the quite strong interaction between the water H atom and the P atom [$d(\mathrm{P-H})=1.79$\,\AA], which lost the chalcogen neighbour. In accordance with the stronger interaction as compared to the adsorption on the pristine monolayer, some changes in the molecular structure can be observed. These are: a significant elongation of the H--O bond from $0.97$\,\AA\ for the gas-phase molecule to $0.98/1.06$\,\AA\  and $0.98/1.04$\,\AA\ for the adsorbed molecule on NiPS$_{3-x}$ and NiPSe$_{3-x}$, respectively. As a result of the interaction with the monolayer, the angle H--O--H is enhanced to $109.8^\circ$ and $109.3^\circ$ vs. $104.5^\circ$ in the case of H$_2$O/NiPS$_{3-x}$ and H$_2$O/NiPSe$_{3-x}$ with respect to the gas phase H$_2$O molecule. 

As a further confirmation of strong interaction between the molecule and the monolayers, we find a strong accumulation of electron density between O and two Ni atoms as well as between H and P atoms (Fig.~\ref{fig:h2o}b). Analysis of the partial DOS (Fig.~\ref{fig:h2o}e) shows that, the orbitals of water contribute significantly to the NiPX$_3$ defect states. Furthermore, the $1b_1$ and $3a_1$ MOs participate in the bonding to the monolayer. (For the DOS of H$_2$O/NiPSe$_{3-x}$, see Supplementary Information, Fig.\,S6). The X-defect formation energy is lowered from $1.48$\,eV ($1.28$\,eV) for the dry monolayer to $1.08$\,eV ($0.82$\,eV) for the surface with an adsorbed water molecule, in case of X = S (Se) respectively. 

In addition to the molecular adsorption structures, dissociated structures were also investigated. In the case of pristine monolayers, all of them are characterised by positive adsorption energies and we do not discuss them here. Yet, the O--H bond elongation observed for the adsorption on defective monolayer, can serve as an indication for the possible dissociation of H$_2$O. Indeed, in the case of NiPS$_{3-x}$ the dissociative adsorption is exothermic process. Consistently with the electronic structure of the defective monolayer, upon molecule dissociation, hydrogen substitutes the P-dangling bond and the OH fragment is bonded to Ni$^{2+}$-ions [$d(\mathrm{Ni-O})=2.04$\,\AA], (Fig.~\ref{fig:h2o}c). The hydroxyl group in the vacancy shows an accumulation in the $1\pi$ orbital and depletion in the $3\sigma$ orbital (Fig.~\ref{fig:h2o}c).  Charge accumulation between P and H is also detected. The observed phenomena can also be seen in partial DOS (Fig.~\ref{fig:h2o}f). Still, the corresponding adsorption energy ($E_\mathrm{ads}=-355$\,meV for NiPS$_3$) is lower in magnitude as compared with the value obtained for molecular adsorption structure, making the dissociative adsorption less favourable. The same trend was observed when studying adsorption of water on CrPX$_3$ monolayers~\cite{Xu:2021ab}. 

All our attempts to find a local minimum corresponding to the dissociative adsorption structure similar to one presented in Figure~\ref{fig:h2o}c were failed, when NiPSe$_3$ is considered. Structure optimisation yields spontaneous recombination of H and OH species into an H$_2$O molecule. Presumably, this is due to the lower electronegativity of Se and bad localisation of the corresponding defect state in the considered DOS (Fig.~\ref{fig:dos}c). Indeed, when considering iron oxides, which are known as effective materials for the water splitting, we noticed that the well-localised defect states in the DOS of respective defective surfaces are separated form the valence band states by ca.\,$300\dots600$\,meV and the dissociative adsorption on these surfaces is strongly favoured (by $0.75\dots1.23$\,eV) over the molecular~\cite{Ovcharenko:2016fu,Mulakaluri:2010vy}.

\section{Conclusions}
\label{S:Conclu}

We present a comprehensive first-principles study of initial steps of adsorption of water on the pristine and defective NiPX$_3$ (X = S, Se) trichalcogenide monolayers, analysing also the underlying electronic mechanisms. On pristine monolayers a molecular adsorption is detected and the results obtained for X = S and Se are almost identical unless the chalcogen vacancy is created. Whereas a well-localised occupied defect state was detected in DOS of NiPS$_3$, that was not revealed in DOS of NiPSe$_3$. This can be a reason why water cannot follow the dissociation pathway in case of X = Se. Although dissociation pathway is possible when X = S, molecular adsorption is energetically more favourable also in this case. Overall, we found that in all cases the adsorption of water has a physisorption character with adsorption energies not exceeding $-650$\,meV. For both NiPX$_3$ under study the X vacancy formation energy becomes lower by ca.\,$400$\,meV if an adsorbed water molecule is present. The comparison of our data with previous results for transition-metal oxides allows us to conclude that dissociative water adsorption is connected with the strong localisation (in space and in energy) of the defect state formed after creation of vacancy. The formation of such state, which is effectively energetically split from the valence band states, can be considered as a descriptor for the consideration of the possible water splitting on the pristine and defected surfaces of transition metal trichalcogenides. Our results are first steps on the consideration of the complicated H$_2$O/MPX$_3$ systems and the presented results can be used as a basis for further theoretical and experimental works.

\section{Experimental Section}

Spin-polarised DFT calculations based on plane-wave basis sets of $500$\,eV cutoff energy were performed with the Vienna \textit{ab initio} simulation package (VASP).~\cite{Kresse:1996kg,Kresse:1994cp,Kresse:1993hw} The Perdew-Burke-Ernzerhof (PBE) exchange-correlation functional~\cite{Perdew:1996ky} was employed. The electron-ion interaction was described within the projector augmented wave (PAW) method~\cite{Blochl:1994fq} with Ni ($3p$, $3d$, $4s$), P ($3s$, $3p$), S ($3s$, $3p$) and Se ($4s$, $4p$) states treated as valence states. The Brillouin-zone integration was performed on $\Gamma$-centred symmetry reduced Monkhorst-Pack meshes using a Gaussian smearing with $\sigma = 0.05$\,eV, except for the calculation of total energies. For these calculations, the tetrahedron method with Bl\"ochl corrections~\cite{Blochl:1994ip} was employed. The $12\times12\times4$ and $24\times24\times1$ k-meshes were used for the studies of bulk and monolayer NiPX$_3$, respectively, and the $12\times12\times1$ k-mesh was used for the $2\times2\times1$ supercells consisting of 4-fold unit monolayers. The DFT+$\,U$ scheme~\cite{Anisimov:1997ep,Dudarev:1998dq} was adopted for the treatment of Ni $3d$ orbitals, with the parameter $U_\mathrm{eff}=U-J$ equal to $6$\,eV. Dispersion interactions were considered adding a $1/r^6$ atom-atom term as parameterised by Grimme (``D2'' parameterisation)~\cite{Grimme:2006fc}. The selected approach yields structural parameters, which are in good agreement with available experimental data.~\cite{Gusmao:2017kz,Yan:2021cn}. Furthermore the correct electronic state (charge-transfer insulator and width of the band gap)~\cite{Yan:2021cn} and magnetic states (zAFM)~\cite{Zhang.2021f5i} are reproduced. For comparison reasons, selected calculations were performed using the HSE06 functional~\cite{Heyd:2005hi}. 

When modelling NiPX$_3$ monolayers, the lattice constant in the lateral plane was set according to the optimised lattice constant of bulk NiPX$_3$ and the positions of P and S were fully relaxed. A vacuum gap was set to approximately $20$\,\AA. During structure optimisation, the convergence criteria for energy and force were set equal to $10^{-5}$\,eV and $10^{-2}$\,eV/\AA, respectively. 

The cleavage energy was calculated as 
\begin{equation}\nonumber
\Delta E_\mathrm{cl} =\left(E_{d_0\rightarrow\infty} - E_0\right)/A,
\label{equ:	cleav}
\end{equation}
where $d_0$ and $E_0$ are, respectively, the van der Waals gap and the total energy of bulk crystal; $A$ is the in-plane area.

To extract the exchange interaction parameters between Ni ions spins, the Heisenberg Hamiltonian was considered
\begin{equation}\nonumber
H=\sum_{\langle i,j \rangle}J_1\vec{S}_i \cdot \vec{S}_j+\sum_{\langle \langle i,j \rangle \rangle}J_2\vec{S}_i \cdot \vec{S}_j+\sum_{\langle \langle \langle i,j \rangle \rangle \rangle}J_3\vec{S}_i \cdot \vec{S}_j,
\end{equation}
where $\vec{S}_i$ is the net spin magnetic moment of the Ni ions at site $i$, three different distance magnetic coupling parameters were estimated, considering one central Ni ions interacted with three nearest neighbouring (NN, $J_1$), six next-nearest neighbouring (2NN, $J_2$), and three third-nearest neighbouring (3NN, $J_3$) Ni ions, respectively. Here, the long-range magnetic exchange parameters ($J$) can be obtained as~\cite{Sivadas:2015gq}
\begin{equation}\nonumber
\begin{aligned}
 &J_1=\frac{E_\mathrm{FM}-E_\mathrm{nAFM}+E_\mathrm{zAFM}-E_\mathrm{sAFM}}{8S^2}\,,\\
 &J_2=\frac{E_\mathrm{FM}+E_\mathrm{nAFM}-(E_\mathrm{zAFM}+E_\mathrm{sAFM})}{16S^2}\,, \\
 &J_3=\frac{E_\mathrm{FM}-E_\mathrm{nAFM}-3(E_\mathrm{zAFM}-E_\mathrm{sAFM})}{24S^2}\,,
 \end{aligned}
 \end{equation}
where $S$ is the calculated magnetic moment of the Ni ion and $E_\mathrm{FM}$, $E_\mathrm{nAFM}$, $E_\mathrm{zAFM}$, $E_\mathrm{sAFM}$ are the total energies in ferromagnetic, N\'eel antiferromagnetic, zigzag antiferromagnetic, and stripy antiferromagnetic configurations, respectively.

To estimate $T_\mathrm{N}$ temperature, Monte Carlo simulations were performed within the Metropolis algorithm with periodic boundary conditions~\cite{Metropolis:1953in}. The three exchange parameters $J_1$, $J_2$ and $J_3$ were used in a series of superlattices $L\times L$ ($L=16,32,64$) containing a large amount of magnetic sites to accurately evaluate the value. Upon the heat capacity $C_v(T) = (\langle E^2\rangle - \langle E\rangle^2)/k_B T^2$ reaching the equilibrium state at a given temperature, the $T_\mathrm{N}$ value can be extracted from the peak of the specific heat profile.

The electrically neutral vacancies were created by removing one or two X atoms from the ($2\times 2$) supercells. Thereby, the distance between repeated vacancies in the nearest-neighbour cells is larger than 10 \AA. The defect formation energy is defined as follows
\begin{equation}\nonumber
\Delta E_\mathrm{def} = \frac{1}{n}\left[E(\mathrm{NiPX}_{3-n})+ n\,\mu_\mathrm{X} - E(\mathrm{NiPX}_3)   \right],
\label{equ:defects}
\end{equation}
where $n$ is a number of defects, $E(\mathrm{NiPX}_{3-n})$ and $E(\mathrm{NiPX}_3)$ are the energies of the 2D NiPX$_3$ with and without vacancy, respectively, $\mu_\mathrm{X}$ is the chemical potential of X atom ($\mu_\mathrm{S}=-4.1279$\,eV and $\mu_\mathrm{Se}=-3.4895$\,eV)~\cite{Yang:2020ex}. 

To study the adsorption of a single molecule, a ($2\times2$) supercell was used with one water molecule added. Adsorption energies were calculated as
\begin{equation}\nonumber
\Delta E_\mathrm{ads} = E(\mathrm{A}/\mathrm{NiPX}_3) - \left[E(\mathrm{A}) + E(\mathrm{NiPX}_3)\right],
\label{equ:adsorption}
\end{equation}
where $E(\mathrm{NiPX}_3)$ and $E(\mathrm{A})$ are the energies of the relaxed isolated 2D $\mathrm{NiPX}_3$ and an adsorbate, and $E(\mathrm{A}/\mathrm{NiPX}_3)$ is the energy of their interacting assembly.

\bigskip
\textbf{Supporting Information} \par 
Supporting Information is available from the Wiley Online Library or from the author.

\bigskip

\textbf{Acknowledgements} \par 
This work was supported by the National Natural Science Foundation of China (Grant No. 21973059). We appreciate the High Performance Computing Center of Shanghai University and Shanghai Engineering Research Center of Intelligent Computing System (No. 19DZ2252600) for providing the computing resources and technical support. Y.D. and E.V. thank the ``ENSEMBLE3 - Centre of Excellence for nanophotonics, advanced materials and novel crystal growth-based technologies'' project \newline(GA No. MAB/2020/14) carried out within the International Research Agendas programme of the Foundation for Polish Science co-financed by the European Union under the European Regional Development Fund and the European Union’s Horizon 2020 research and innovation programme Teaming for Excellence (GA. No. 857543) for support of this work. 

\medskip

\bibliographystyle{MSP}

\clearpage

\begin{table}[ht]
\caption{\label{tab:2D}Ground-state properties as obtained for the different magnetic states of 2D NiPS$_3$ and 2D NiPSe$_3$ with PBE$+U$+D2: $E$ (in eV) is the total energy, $\Delta E$ (in meV per (2$\times$1) supercell) is the difference between the energy calculated for the magnetic states and the energy calculated for the zAFM state, $M$ is magnetic moment (in $\mu_B$ per Ni ion), $E_g$ (in eV) is the band gap.
}  
  \begin{tabular}{@{}lllrcl@{}}
 \hline
System &  State & $E$ & $\Delta E$ & $M$ &$E_g$   \\
    \hline
NiPS$_3$	&nAFM	&$-89.443$	&$13$	&$1.59$	&$2.26$			\\
		&zAFM	&$-89.456$	&$0$		&$1.59$	&$2.19$			\\
		&sAFM	&$-89.337$	&$119$	&$1.60$	&$1.69$			\\
		&FM		  &$-89.357$	&$99$	&$1.60$	&$2.34$/$2.21$\textsuperscript{\emph{a}}	\\
		&NM  	 &$-83.321$	&$6135$	&$0$		&$0.00$			\\[0.3cm] 
NiPSe$_3$&nAFM 	&$-80.109$	&$24$ 	&$1.54$	&$1.89$                	 \\
		&zAFM 	&$-80.133$	&$0$         &$1.53$&$1.85$               \\
		&sAFM 	&$-79.996$	&$137$ 	&$1.56$	&$1.32$                	 \\
		&FM   	  &$-80.031$	&$102$ 	&$1.56$	&$1.35$/$1.78$\textsuperscript{\emph{a}}  \\
		&NM   	 &$-74.835$	&$5298$	&$0$		&$0.00$                   	   \\
\hline
  \end{tabular}
  
  \textsuperscript{\emph{a}} two values for the spin-up and spin-down channels are indicated
\end{table}

\clearpage
\begin{table}[ht]
\caption{\label{tab:J} Results for the pristine NiPX$_3$ monolayer obtained for the zAFM state: $a$ (in \AA) is the in-plane lattice constant, $h$ (in \AA) is the monolayerlayer thickness;  $d_\mathrm{Ni-X}$, $d_\mathrm{P-X}$, $d_\mathrm{P-P}$ (in \AA) and $\theta_\mathrm{Ni-X-Ni}$ (in degrees) are the respective distances and angles; ($J$) in meV are the exchange-coupling parameters; $T_\mathrm{N}$ (in K) is N\'eel temperature. For comparison available experimental data are also presented.}
  \begin{tabular}{@{}llllllllllll@{}}
 \hline
X	&Method	&$a$		&$h$		&$d_\mathrm{Ni-X}$&$d_\mathrm{P-X}$&$d_\mathrm{P-P}$&$\theta_\mathrm{Ni-X-Ni}$& $J_1$&$J_2$&$J_3$&$T_\mathrm{N}$ \\
\hline
S
&Calc.\textsuperscript{\emph{a}}	&$5.85$ &$3.13$ 	&$2.49$ 	&$2.04$ 	&$2.18$ 	&$85.3$ 	
&$1.503$	&$0.098$	&$-7.316$	&$148$\\
&Calc.\textsuperscript{\emph{b}}	&$5.82$ &		 	&$2.44$ 	&$2.05$ 	&$2.18$ 	&$87.1$
&	&	&	&  \\
&Expt.&$5.82$\textsuperscript{\emph{c}}	&		&$2.50$\textsuperscript{\emph{c}}	&$1.98$\textsuperscript{\emph{c}}	&$2.17$\textsuperscript{\emph{c}}	&$84.4$\textsuperscript{\emph{c}}	
&$1.84$\textsuperscript{\emph{d}}&$-0.18$\textsuperscript{\emph{d}}&$-6.95$\textsuperscript{\emph{d}}&$155$\textsuperscript{\emph{d,e}}
\\[0.3cm] 
Se
&Calc.\textsuperscript{\emph{a}}	&$6.16$& $3.30$&$2.60$&$2.21$&$2.20$&$86.3$
&$3.064$	&$0.303$	&$-8.509$&$183$ \\
&Calc.\textsuperscript{\emph{b}}	&$6.17$ &		 	&$2.54$ 	&$3.45$ 	&$2.22$ 	&$89.1$
&	&	&	&  \\
&Expt.&$6.13$\textsuperscript{\emph{f}}	&		&$2.61$\textsuperscript{\emph{f}}	&$2.09$\textsuperscript{\emph{f}}	&$2.24$\textsuperscript{\emph{f}}	&$85.2$\textsuperscript{\emph{f}}	
&		&		&		&$206$\textsuperscript{\emph{e}} \\[0.15cm] 
 \hline

  \end{tabular}
 
\textsuperscript{\emph{a}} This work;
\textsuperscript{\emph{b}} Ref.~\cite{Gu:2019kw};
\textsuperscript{\emph{c}} Ref.~\cite{Brec:1980ts};
\textsuperscript{\emph{d}} Ref.~\cite{Lancon:2018gf};
\textsuperscript{\emph{e}} Ref.~\cite{LeFlem:1982gs};
\textsuperscript{\emph{f}} Ref.~\cite{Rao:1991bj}

\end{table}

\clearpage
\begin{table}[ht]
\caption{Results for the NiPX$_{3-x}$ monolayer: $\Delta E_\mathrm{def}$ (in eV) is defect formation energy; $d_\mathrm{Ni-X}$, $d_\mathrm{P-X}$, $d_\mathrm{P-P}$ (in \AA) and $\theta_\mathrm{Ni-X-Ni}$ (in degrees) are the respective distances and angles.
}
  \label{tab:df}
  \begin{tabular}{@{}lllllll@{}}
  \hline
	 X&Vacancy type &$\Delta E_\mathrm{def}$&$d_\mathrm{Ni-X} $&$d_\mathrm{P-X}$&$d_\mathrm{P-P}$&$\theta_\mathrm{Ni-X-Ni}$\\
	 \hline
      S&V$_\mathrm{S}$@1L	&  $1.480$    &$2.45/2.50$ &$2.10$    &$2.17$  &$86.3/86.4$ \\
	    &V$_\mathrm{S2}$@1L& $1.632$    &$2.47$          &$2.11$    &$2.12$  &$86.3$  \\
	    &V$_\mathrm{S2}$@2L&$1.560$   &$2.39/2.47$&$2.11$    &$2.17$      &$88.2/88.3$	 \\[0.3cm]

	Se&V$_\mathrm{Se}$@1L&$1.279$    &$2.60/2.61$ &$2.27$  &$2.18$& $86.3/86.3$   \\
	    &V$_\mathrm{Se2}$@1L& $1.547$ &$2.61/2.62$ &$2.29$  &$2.11$  &$85.6$  \\
	    &V$_\mathrm{Se2}$@2L&$1.361$&$2.54/2.59$ &$2.28$ &$2.18$ &$87.7/87.8$	 \\
\hline
  \end{tabular}

 \end{table}

\clearpage

\clearpage

\clearpage

\end{document}